\documentclass[runningheads]{llncs}
\usepackage{amsmath}
\usepackage{amssymb}
\usepackage{amsthm}
\usepackage{setspace,xcolor}
\usepackage{hyperref}

\usepackage{xspace}
\usepackage{comment}
\usepackage{mdframed}

\usepackage{paralist}
\usepackage[ruled, noend]{algorithm2e}

\usepackage[shortlabels]{enumitem}
\usepackage{cleveref}

\newtheorem{thm}{Theorem}[section]
\newtheorem{prop}[thm]{Proposition}
\newtheorem{lem}[thm]{Lemma}

\def\Opt{\mbox{\rm Opt}\xspace}
\def\lowlax{\mathcal{LL}\xspace}
\def\highlax{\mathcal{HL}\xspace}

\def\inviable{\mathcal{N}\xspace}

\newcommand{\LMNYalg}{\textsc{LMNY}\xspace}
\newcommand{\laxalg}{\textsc{Lax}\xspace}
\newcommand{\srptalg}{\textsc{SRPT}\xspace}
\newcommand{\mlaxalg}{\textsc{MLax}\xspace}
\newcommand{\finalalg}{\textsc{FinalAlg}\xspace}

\begin{document}

\title{A Competitive Algorithm for Throughput Maximization on Identical Machines}
\titlerunning{Throughput Maximization on Identical Machines}
%
\author{Benjamin Moseley\thanks{B. Moseley and R. Zhou were supported in part by NSF grants  CCF-1824303, CCF-1845146, CCF-1733873 and CMMI-1938909.  Benjamin Moseley was additionally supported in part by a Google Research Award, an Infor Research Award, and a Carnegie Bosch Junior Faculty Chair.}\inst{1} \and
Kirk Pruhs\thanks{Supported in part  by NSF grants   CCF-1907673,  CCF-2036077 and an IBM Faculty Award.}\inst{2}\orcidID{0000-0001-5680-1753} \and
Clifford Stein\thanks{Research partly supported by NSF Grants CCF-1714818 and CCF-1822809.}\inst{3} \and
Rudy Zhou\inst{1} }
\authorrunning{B. Moseley et al.}
%
\institute{Tepper School of Business, Carnegie Mellon University \email{\{moseleyb,rbz\}@andrew.cmu.edu} \url{https://www.andrew.cmu.edu/user/moseleyb/} \url{https://rudyzhou.github.io/} \and
Computer Science Department, University of Pittsburgh
\email{kirk@cs.pitt.edu}\\
\url{https://people.cs.pitt.edu/~kirk/} \and
Industrial Engineering and Operations Research, Columbia University \email{cliff@ieor.columbia.edu}}
\maketitle              
\begin{abstract}
This paper considers the basic problem of scheduling  jobs online with preemption to maximize the number of jobs completed by their deadline on $m$ identical machines. The main result is an $O(1)$ competitive deterministic algorithm for any number of machines $m >1$.
\end{abstract}

\section{Introduction}\label{sec_intro}

We consider the basic problem of preemptively scheduling jobs that arrive online with sizes and deadlines on $m$ identical machines to maximize the number of jobs that complete by their deadline.

\begin{definition}[Throughput Maximization]\label{def_problem}
    Let $J$ be a collection of jobs such that each $j \in J$ has a \emph{release time} $r_j$, a \emph{processing time} (or \emph{size}) $x_j$, and a \emph{deadline} $d_j$. The jobs arrive online at their release times, at which the scheduler becomes aware of job $j$ and its $x_j$ and $d_j$.
    
    At each moment of time, the scheduler
    can specify up to $m$ released jobs to run
    at this time, and the remaining processing time
    of the  jobs that are run is decreased at a unit rate (so we assume that the online scheduler is allowed to produce a migratory schedule.) A job is completed if its remaining processing time drops to zero by the deadline of that job.
    The objective is to maximize the number completed jobs.

\end{definition}

A key concept in designing algorithms for this problem is the laxity of a job. The \emph{laxity} of a job $j$ is $\ell_j = (d_j - r_j) - x_j$, which is the maximum amount of time we can not run $j$ and still possibly complete it. 

We measure the performance of our algorithm by the \emph{competitive ratio}, which is the maximum over all instances of the ratio of the objective value of our algorithm to the objective value of the optimal offline schedule that is aware of all jobs in advance.

This problem is well understood for the $m=1$ machine case.
No $O(1)$-competitive deterministic algorithm is possible~\cite{BaruahKMMRRSW92},
but there is a randomized algorithm that is 
$O(1)$-competitive against an oblivious adversary~\cite{KalyanasundaramP03},
and there is a scalable ($O(1+\epsilon)$-speed $O(1/\epsilon)$-competitive) deterministic algorithm~\cite{KalyanasundaramP00}. The scalability result in \cite{KalyanasundaramP00}
was extended  to the case of $m > 1 $ machines in \cite{LucierMNY13}. 

Whether an $O(1)$-competitive algorithm exists for $m > 1$ 
machines has been open for twenty years. Previous results for the multiple machines setting require resource augmentation or assume that all jobs have high laxity \cite{LucierMNY13,EberleMS20}.

The main issue issue in removing these assumptions is
how to determine which machine
to assign a job to. If an online algorithm could determine which
machine each job was assigned to in \Opt, we could obtain an $O(1)$-competitive algorithm
for $m > 1$ machines by a relatively straight-forward
adaptation of the results from \cite{KalyanasundaramP03}. 
However, if the online algorithm ends up assigning some jobs to different
machines than \Opt, then 
comparing the number of completed jobs is challenging. Further, if jobs have small laxity, then the algorithm can be severely penalized for small mistakes in this assignment.  
One way to view the speed augmentation (or high laxity assumption) analyses in \cite{LucierMNY13,EberleMS20} is that the speed augmentation assumption allows one to avoid having to address this issue in the analyses.


\subsection{Our Results}

Our main result is an  $O(1)$-competitive deterministic algorithm
for Throughput Maximization on $m>1$ machines.


\begin{thm}\label{thm_main_deterministic}
    For all $m > 1$, there exists a deterministic $O(1)$-competitive algorithm for Throughput Maximization on $m$ machines.
\end{thm}

We summarize our results and prior work in Table \ref{table}. Interestingly, notice that on a single machine there is no constant competitive deterministic algorithm, yet a randomized algorithm exists with constant competitive ratio.  Our work shows that once more than one machine is considered, then determinism is sufficient to get a $O(1)$-competitive online algorithm. 

\begin{table}[h]
\begin{tabular}{|l|l|l|l|}
\hline
            & Deterministic & Randomized & \begin{tabular}[c]{@{}l@{}}Speed\\ Augmentation\end{tabular} \\ \hline
     \hline
     $m=1$       & $\omega(1)$   & $O(1)$    & $O(1+\epsilon)$-speed $O(1/\epsilon)$-competitive        \\ 
          &\cite{BaruahKMMRRSW92}   & \cite{KalyanasundaramP03}    & \cite{KalyanasundaramP00}        \\ \hline
 $m > 1$ &        $O(1)$       &      $O(1)$       & $O(1+\epsilon)$-speed $O(1/\epsilon)$-competitive        \\ 
 &           [This paper]    &     [This paper]       &  \cite{LucierMNY13}        \\  
 \hline
\end{tabular}
\medskip
\caption{Competitiveness Results}
\label{table}
\end{table}

\subsection{Scheduling Policies}

We give some basic definitions and notations about scheduling policies.

A job $j$ is \emph{feasible} at time $t$ (with respect to some schedule) if it can still be feasibly completed, so $x_j(t) > 0$ and $t + x_j(t) \leq d_j$, where $x_j(t)$ is the remaining processing time of job $j$ at time $t$ (with respect to the same schedule.)

Then a schedule $\mathcal{S}$ of jobs $J$ is defined by a map from
time/machine pairs $(t, i)$ to a feasible job $j$ 
that is run on machine $i$ at time
$t$, with the constraint that no job can be
run one two different  machines at the same time. . We conflate $\mathcal{S}$ with the scheduling policy as well as the set of jobs \emph{completed} by the schedule. Thus, the objective value achieved by this schedule is $\lvert \mathcal{S} \rvert$.
    
A schedule is \emph{non-migratory} if for every job $j$  there exists a 
machine $i$ such that if $j$ is run at time $t$ then $j$ is run on machine $i$. Otherwise the schedule is \emph{migratory}.

If $\mathcal{S}$ is a scheduling algorithm, then $\mathcal{S}(J,m)$
denotes the schedule that results from running $\mathcal S$
on instance $J$ on $m$ machines.
Similarly, $\Opt(J,m)$ denotes the optimal schedule on instance $J$ on $m$ machines. We will sometimes omit the $J$ and/or the $m$ if 
they are clear from context. Sometimes we will abuse notation and let $\Opt$ denote a nearly-optimal schedule 
that additionally has some desirable structural property.

\subsection{Algorithms and Technical Overview}

A simple consequence of the results in
\cite{KalyanasundaramP01} and \cite{KalyanasundaramP03} is
an $O(1)$-competitive algorithm in the case that $m =O(1)$.
Thus we concentrate on the case that $m$ is large. 
Also note that since there is an $O(1)$-approximate
non-migratory schedule~\cite{KalyanasundaramP01}, 
changing the number of machines by an $O(1)$ factor
does not change the optimal objective value by 
more than an $O(1)$ factor. 
This is because we can always take an optimal non-migratory schedule on $m$ machines and create a new schedule on $m/c$  machines 
whose objective value decreases by at most a factor of $c$,
by keeping the $m/c$ machines that complete the most jobs.

These observations about the structure of near-optimal
schedules allow us to design a
$O(1)$-competitive algorithm that is a combination of various
deterministic algorithms. In particular, on instance $J$ our algorithm, \finalalg  will run a deterministic algorithm   \LMNYalg
on $m/3$ machines on the
subinstance $J_{hi} = \{j \in J \mid \ell_j > x_j\}$  of high laxity jobs,
a deterministic algorithm \srptalg
on $m/3$ machines on the subinstance $J_{lo} = \{j \in J \mid \ell_j \leq x_j\}$
   of low laxity jobs, and a deterministic algorithm \mlaxalg
on $m/3$ machines on the subinstance $J_{lo}$
   of low laxity jobs. Note that we run \srptalg and \mlaxalg on the same jobs. To achieve this, if both algorithms decide to run the same job $j$, then the algorithm in which $j$ has shorter remaining processing time actually runs job $j$, and the other simulates running $j$.
   
We will eventually show that
for all instances, at least one of these three algorithms
is $O(1)$-competitive, from which our main result will follow. Roughly, each of the three algorithms is responsible for a different part of \Opt.

Our main theorem about \finalalg is the following:

\begin{thm}\label{thm_main_final}
    For any $m \geq 48$, \finalalg is a $O(1)$-competitive deterministic algorithm for Throughput Maximization on $m$ machines.
\end{thm}

We now discuss these
three component algorithms of \finalalg.

\subsubsection{\LMNYalg}

The algorithm \LMNYalg is the algorithm from \cite{LucierMNY13}
with the following guarantee.

\begin{lem}\label{lem_LMNY}\cite{LucierMNY13}
    For any number of machines $m$, and any job instance $J$,
    \LMNYalg is an $O(1)$-competitive deterministic algorithm 
    on the instance $J_{hi}$.
\end{lem}


\subsubsection{\srptalg}

The algorithm   \srptalg   
is a variation of the standard shortest remaining processing time algorithm:

\begin{definition}[\srptalg]\label{def_srpt}
    At each time, run the $m$ feasible jobs with shortest remaining processing time. If there are less than $m$ feasible jobs, then all feasible jobs are run. 
\end{definition}

We will show that \srptalg is competitive with the low laxity jobs in that are not preempted in \Opt.

\subsubsection{\mlaxalg}

The final, most challenging, component algorithm of \finalalg is \mlaxalg, which intuitively we want to be competitive on  low-laxity jobs 
in \Opt that are preempted.

    To better understand the challenge
    of achieving this, consider $m=1$ and 
    an instance of disagreeable jobs,
    which means that jobs with an earlier release time have a later
    deadline. Further, suppose all jobs but one in \Opt is preempted and completed at a later time.

     To be competitive, \mlaxalg must preempt almost all
     the jobs that it completes, but cannot afford to abandon
     too many jobs that it preempts. Because the
     jobs have low laxity, this can be challenging as 
     it can only preempt each job for a small amount of time,
     and its hard to know which of the many options is
     the ``right'' job to preempt for. 
     This issue was resolved in \cite{KalyanasundaramP03} 
     for the case of $m=1$ machine, but the issue gets more challenging when $m > 1$, because we also have to choose the ``right'' machine to assign a job.

We now describe the algorithm \mlaxalg. 
Let $\alpha = O(1)$ be a sufficiently large constant (chosen later.) \mlaxalg maintains $m$ stacks (last-in-first-out data structures) of jobs (one per machine), $H_1, \dots, H_m$. The stacks are initially empty. At all times, \mlaxalg runs the top job of stack $H_i$ on machine $i$. We define the \emph{frontier} $F$ to be the set consisting of the top job of each stack (i.e. all currently running jobs.) It remains to describe how the $H_i$'s are updated.

There are two types of events that cause \mlaxalg to update the $H_i$'s: reaching a job's pseudo-release time (defined below) or completing a job.

\begin{definition}[Viable Jobs and Pseudo-Release Time]\label{def_viable}
    The pseudo-release time (if it exists) $\tilde{r}_j$ of job $j$ is the earliest time in $[r_j, r_j + \frac{\ell_j}{2}]$ such that there are at least $\frac{7}{8}m$ jobs $j'$ on the frontier satisfying $\alpha x_{j'} \geq \ell_j$.
    
    We say a job $j$ is \emph{viable} if $\tilde{r}_j$ exists and \emph{non-viable} otherwise.
\end{definition}

At job $j$'s pseudo-release time (note $\tilde{r}_j$ can be determined online by \mlaxalg), \mlaxalg does the following:
\begin{enumerate}[a)]
    \item If there exists a stack whose top job $j'$ satisfies $\alpha x_j \leq \ell_{j'}$, then \emph{push} $j$ onto any such stack.\label{line_push}
    \item Else if there exist at least $\frac{3}{4} m$ stacks whose second-top job $j''$ satisfies $\alpha x_j \leq \ell_{j''}$ and further some such stack has top job $j'$ satisfying $\ell_j > \ell_{j'}$, then on such a stack with minimum $\ell_{j'}$, \emph{replace} its top job $j'$ by $j$. \label{line_replace} 
\end{enumerate}
While the replacement operation in step \ref{line_replace} can be implemented as a pop and then push, we view it as a separate operation for analysis purposes.
To handle corner cases in these descriptions, one can assume that there is a job with infinite size/laxity on the  bottom of each $H_i$.

When \mlaxalg completes a job $j$ that was on stack $H_i$, \mlaxalg does the following:
\begin{enumerate}
    \item[c)] \emph{Pop} $j$ off of stack $H_i$.\label{line_completion_pop}
    \item[d)] Keep \emph{popping} $H_i$ until the top job of $H_i$ is feasible.\label{line_infeasible_pop}
\end{enumerate}

\subsubsection{Analysis Sketch}


There are three main steps in proving \Cref{thm_main_final} to show \finalalg is $O(1)$-competitive:
\begin{itemize}
    \item In \Cref{sec_prelim}, we show how to modify the optimal schedule to obtain certain structural properties that facilitate the comparison with \srptalg and \mlaxalg.
    \item In \Cref{sec_nonviable}, we show that \srptalg is competitive with the low-laxity, non-viable jobs.
    Intuitively, the jobs that \mlaxalg is running that prevent
    a job $j$ from becoming viable are so much smaller than 
    job $j$, and they provide a witness that \srptalg must also be working on jobs much smaller than $j$.
    \item In \Cref{sec_viable}, we show that \srptalg and \mlaxalg together are competitive with the low-laxity, viable jobs.
    First, we show that \srptalg is competitive with the number of
    non-preempted jobs in \Opt. We then essentially show that \mlaxalg is
    competitive with the number of preempted jobs in \Opt. 
    The key component is the design of \mlaxalg is the condition
    that a job $j$ won't replace a job on the frontier unless
    at there are at least $\frac{3}{4} m$ stacks whose second-top job $j''$ satisfies $\alpha x_j \leq \ell_{j''}$. This is the condition that intuitively most differentiates the \mlaxalg from
    $m$ copies of the \laxalg algorithm in \cite{KalyanasundaramP03}. This also is the
    condition that allows us to surmount the issue of potentially assigning a job to a ``wrong'' processor. Jobs that satisfy this condition are highly flexible about where they can go on the frontier. Morally, our analysis shows that a constant fraction of the jobs that \Opt preempts and completes must be such flexible jobs. 
\end{itemize}

We combine these results in \Cref{sec_final} to complete the analysis of \finalalg.

    If $S$ is a schedule,
    let $|S|$ be the number of jobs completed in $S$.
    If $\rm A$ is a scheduling algorithm, we let
    ${\rm A}(\mathcal{K})$ denote the schedule produced by ${\rm A}$ on
    input $\mathcal{K}$. We let $\Opt(\mathcal{K})$ denote the optimal
    schedule on an instance $\mathcal{K}$.

    A schedule $S$ is {\em resolute} if $S$ completes every job
    that it runs. 
    We say that $S \subseteq T$ if every job completed in
    $S$ is completed in $T$.
    If $S$ runs a job $J_i$ then we denote the last time that $J_i$ is
    run as
    $c_i(S)$. 
    If $S$ completes a job $J_i$ then $c_i(S)$ is the {\em completion
    time} of $J_i$.
    If $S$ runs a job $J_i$ then we denote the first time that $J_i$
    is run as
    $f_i(S)$. 
    A job $J_i$ completed in a schedule $S$ is {\em idle}
    if the machine is idle for some non-zero amount of time between when $J_i$
    is first run and when $J_i$ is completed.
    We say a schedule $S$ is {\em efficient} if
    whenever there is a job that may still complete by its deadline in $S$, 
    $S$ is running some such job, that is, $S$ never unnecessarily idles
    the machine.
    
     We say a single-machine schedule $\mathcal{S}$ is a forest schedule if for all jobs $j,j'$ such that $f_j(\mathcal{S}) < f_{j'}(\mathcal{S})$, then $\mathcal{S}$ does not run $j$ during the time interval $(f_{j'}(\mathcal{S}), c_{j'}(\mathcal{S}))$ (so the $(f_j(\mathcal{S}), c_j(\mathcal{S}))$-intervals form a laminar family.) Then $\mathcal{S}$ naturally defines a forest (in the graph-theoretic sense), where the nodes are jobs run by $\mathcal{S}$ and the descendants of a job $j$ are the the jobs that are first run in the time interval $(f_j(\mathcal{S}), c_j(\mathcal{S}))$.
    A non-migratory $m$-machine schedule is a forest schedule if all of its single-machine schedules are forest schedules.
    Because the Earliest Deadline First (EDF) algorithm 
    feasible completes all jobs on one machine if there
    is a feasible schedule on one machine, the optimal non-migratory
    schedule is a forest schedule. 

\end{comment}

\subsection{Related Work}


There is a line of papers that consider a dual 
version of the problem, where there is a constraint that all jobs
must be completed by their deadline, and the objective is to minimize
the number of machines used~\cite{PhillipsSTW02,CMS18,AzarC18,ImMPS17}.
The current best known bound on the competitive ratio for this version is $O(\log \log m)$
from \cite{ImMPS17}. 

The speed augmentation results in \cite{KalyanasundaramP00,LucierMNY13} for throughput can be generalized to weighted 
throughput, where there a profit for each job, and the objective is
to maximize the aggregate profit of jobs completed by their deadline. 
But without speed augmentation, $O(1)$-approximation is not possible
for weighted throughput for any $m$, even allowing
randomization~\cite{KorenS94}.

There is also a line of papers that consider variations on 
online throughput scheduling in which the
online scheduler has to commit to completing jobs at some point in time,
with there being different variations of when commitment is required~\cite{LucierMNY13,EberleMS20,ChenEMSS20}.
For example, \cite{EberleMS20} showed that there is a scalable
algorithm for online throughput maximization that commits to finishing 
every job that it begins executing. 

\section{Structure of Optimal Schedule}\label{sec_prelim}

The goal of this section is to introduce the key properties of (near-)optimal scheduling policies that we will use in our analysis.

For completeness, we show that by losing a constant factor in the competitive ratio, we can use a constant factor fewer machines than $\Opt$. This justifies \finalalg running each of three algorithms on $\frac{m}{3}$ machines.

\begin{lem}\label{lem_reduce_machine}
    For any collection of jobs $J$, number of machines $m$, and $c  >1$, we have $\lvert \Opt(J, \frac{m}{c}) \rvert = \Omega( \frac{1}{c}\lvert \Opt(J, m) \rvert)$.
\end{lem}
\begin{proof}
    It is shown in \cite{KalyanasundaramP01} that for any schedule on $m$ machines, there exists a non-migratory schedule on at most $6m$ machines that completes the same jobs. Applied to $\Opt(J,m)$, we obtain a non-migratory schedule $\mathcal{S}$ on $6m$ machines with $\lvert \mathcal{S} \rvert = \lvert \Opt(J,m) \rvert$. Keeping the $\frac{m}{c}$ machines that complete the most jobs in $\mathcal{S}$ gives a non-migratory schedule on $\frac{m}{c}$ machines that completes at least $\frac{1}{6c}\lvert \mathcal{S} \rvert$ jobs.
\end{proof}

A non-migratory schedule on $m$ machines can be expressed as $m$ schedules, each on a single machine and a separate set of jobs. To characterize these single machine schedules, we introduce the concept of forest schedules. Let $\mathcal{S}$ be any schedule. For any job $j$, we let $f_j(\mathcal{S})$ and $c_j(\mathcal{S})$ denote the first and last times that $\mathcal{S}$ runs the job $j$, respectively. Note that $\mathcal{S}$ does not necessarily complete $j$ at time $c_j(\mathcal{S})$.

\begin{definition}[Forest Schedule]\label{def_forest_sched}
    We say a single-machine schedule $\mathcal{S}$ is a forest schedule if for all jobs $j,j'$ such that $f_j(\mathcal{S}) < f_{j'}(\mathcal{S})$, then $\mathcal{S}$ does not run $j$ during the time interval $(f_{j'}(\mathcal{S}), c_{j'}(\mathcal{S}))$ (so the $(f_j(\mathcal{S}), c_j(\mathcal{S}))$-intervals form a laminar family.) Then $\mathcal{S}$ naturally defines a forest (in the graph-theoretic sense), where the nodes are jobs run by $\mathcal{S}$ and the descendants of a job $j$ are the the jobs that are first run in the time interval $(f_j(\mathcal{S}), c_j(\mathcal{S}))$.
    
    Then a non-migratory $m$-machine schedule is a forest schedule if all of its single-machine schedules are forest schedules.
\end{definition}

With these definitions, we are ready to construct the near-optimal policies that we will compare \srptalg and \mlaxalg to:

\begin{lem}\label{lem_opt_struct}
    Let $J$ be a set of jobs satisfying $\ell_j \leq x_j$ for all $j \in J$. Then for any times $\hat{r}_j \in [r_j, r_j + \frac{\ell_j}{2}]$ and constant $\alpha \geq 1$, there exist non-migratory forest schedules $\mathcal{S}$ and $\mathcal{S}'$ on the jobs $J$ such that:
    \begin{enumerate}
        \item Both $\mathcal{S}$ and $\mathcal{S}'$ complete every job they run.\label{prop_resolute}
        \item Let $J_i$ be the set of jobs that $\mathcal{S}$ runs on machine $i$. For every machine $i$ and time, if there exists a feasible job in $J_i$, then $\mathcal{S}$ runs such a job.\label{prop_efficient}
        \item For all jobs $j \in \mathcal{S}$, we have $f_j(\mathcal{S}) = \hat{r}_j$.\label{prop_start}
        \item If job $j'$ is a descendant of job $j$ in $\mathcal{S}$, then $\alpha x_{j'} \leq \ell_j$\label{prop_shrinking}
        \item $\lvert \{\text{leaves of $\mathcal{S}'$}\}\rvert + \lvert \mathcal{S} \rvert = \Omega(\lvert \Opt(J) \rvert)$.\label{prop_leafy}
    \end{enumerate}
\end{lem}
\begin{proof}
    We modify the optimal schedule $\Opt(J)$ to obtain the desired properties. First, we may assume that $\Opt(J)$ is non-migratory by losing a constant factor (\Cref{lem_reduce_machine}.) Thus, it suffices to prove the lemma for a single machine schedule, because we can apply the lemma to each of the single-machine schedules in the non-migratory schedule $\Opt(J)$. The proof for the single-machine case follows from the modifications given in Lemmas 22 and 23 of \cite{KalyanasundaramP03}. We note that \cite{KalyanasundaramP03} only show how to ensure $f_j(\mathcal{S}) = t_j$ for a particular $t_j \in [r_j, r_j + \frac{\ell_j}{2}]$, but it is straightforward to verify that the same proof holds for any $t_j \in [r_j, r_j + \frac{\ell_j}{2}]$.
\end{proof}

Morally, the schedule $\mathcal{S}$ captures the jobs in \Opt that are preempted and $\mathcal{S'}$ captures the jobs in \Opt that are not preempted (i.e. the leaves in the forest schedule.)

\section{\srptalg is Competitive with Non-Viable Jobs}\label{sec_nonviable}

The main result of this section is that \srptalg is competitive with the number of non-viable, low-laxity jobs of the optimal schedule (\Cref{thm_srpt_nv_low}.) We recall that a job $j$ is non-viable if for \emph{every time} in $[r_j, r_j + \frac{\ell_j}{2}]$, there are at least $\frac{1}{8}m$ jobs $j'$ on the frontier of \mlaxalg satisfying $\alpha x_{j'} < \ell_j$.

\begin{thm}\label{thm_srpt_nv_low}
    Let $J$ be a set of jobs satisfying $\ell_j \leq x_j$ for all $j \in J$. Then for $\alpha = O(1)$ sufficiently large and number of machines $m \geq 16$, we have $\lvert \srptalg(J) \rvert = \Omega( \lvert \Opt(J_{nv}) \rvert)$, where $J_{nv}$ is the set of non-viable jobs with respect to $\mlaxalg(J)$.
\end{thm}

In the remainder of this section, we prove \Cref{thm_srpt_nv_low}. The main idea of the proof is that for any non-viable job $j$, \mlaxalg is running many jobs that are much smaller than $j$ (by at least an $\alpha$-factor.) These jobs give a witness that \srptalg must be working on these jobs or even smaller ones.

We begin with a lemma stating that \srptalg is competitive with the leaves of any forest schedule. Intuitively this follows because whenever some schedule is running a feasible job, then \srptalg~ either runs the same job or a job with shorter remaining processing time. We will use this lemma to handle the non-viable jobs that are not preempted.

\begin{lem}\label{lem_srpt_leaves}
    Let $J$ be any set of jobs and $\mathcal{S}$ be any forest schedule on $m$ machines and jobs $J' \subset J$ that only runs feasible jobs. Let $L$ be the set of leaves of $\mathcal{S}$. Then $\lvert \srptalg(J) \rvert \geq \frac{1}{2} \lvert L \rvert$
\end{lem}
\begin{proof}
    It suffices to show that $\lvert L \setminus \srptalg(J) \rvert \leq \lvert \srptalg(J) \rvert$. The main property of \srptalg~ gives:
    
    \begin{prop}\label{prop_srpt_leaves}
        Consider any leaf $\ell \in L \setminus \srptalg(J)$. Suppose $\mathcal{S}$ starts running $\ell$ at time $t$. Then \srptalg~ completes $m$ jobs in the time interval $[f_\ell(\mathcal{S}), f_\ell(\mathcal{S}) + x_\ell]$.
    \end{prop}
    \begin{proof}
        At time $f_\ell(\mathcal{S})$ in \srptalg(J), job $\ell$ has remaining processing time at most $x_\ell$ and is feasible by assumption. Because $\ell \notin \srptalg(J)$, there must exist a first time $t' \in [f_\ell(\mathcal{S}), f_\ell(\mathcal{S}) + x_\ell]$ where $\ell$ is not run by $\srptalg(J)$. At this time, $\srptalg(J)$ must be running $m$ jobs with remaining processing time at most $x_\ell - (t' - f_\ell(\mathcal{S}))$. In particular, $\srptalg(J)$ must complete $m$ jobs by time $f_\ell(\mathcal{S}) + x_\ell$.
    \end{proof}
    
    Using the proposition, we give a charging scheme: Each job $\ell \in L \setminus \srptalg(J)$ begins with $1$ credit. By the proposition, we can find $m$ jobs that $\srptalg(J)$ completes in the time interval $[f_\ell(\mathcal{S}), f_\ell(\mathcal{S}) + x_\ell]$. Then $\ell$ transfers $\frac{1}{m}$ credits each to $m$ such jobs in \srptalg.
    
    It remains to show that each $j \in \srptalg(J)$ gets at most $1$ credit. Note that $j$ can only get credits from leaves $\ell$ such that $c_j(\srptalg) \in [f_\ell(\mathcal{S}), f_\ell(\mathcal{S}) + x_\ell]$. There are at most $m$ such intervals (at most one per machine), because we only consider leaves, whose intervals are disjoint if there are on the same machine. 
\end{proof}

Now we are ready to prove \Cref{thm_srpt_nv_low}.

\begin{proof}[Proof of \Cref{thm_srpt_nv_low}]
    Let $\mathcal{S}, \mathcal{S}'$ be the schedules guaranteed by \Cref{lem_opt_struct} on the set of jobs $J_{nv}$ with $\hat{r}_j = r_j$ for all $j \in J_{nv}$. We re-state the properties of these schedules for convenience:
    \begin{enumerate}
        \item Both $\mathcal{S}$ and $\mathcal{S}'$ complete every job they run.
        \item Let $J_i$ be the set of jobs that $\mathcal{S}$ runs on machine $i$. For every machine $i$ and time, if there exists a feasible job in $J_i$, then $\mathcal{S}$ runs such a job.
        \item For all jobs $j \in \mathcal{S}$, we have $f_j(\mathcal{S}) = r_j$.
        \item If job $j'$ is a descendant of job $j$ in $\mathcal{S}$, then $\alpha x_{j'} \leq \ell_j$
        \item $\lvert \{\text{leaves of $\mathcal{S}'$}\}\rvert + \lvert \mathcal{S} \rvert = \Omega(\lvert \Opt(J_{nv}) \rvert)$.
    \end{enumerate}
    By \Cref{lem_srpt_leaves}, we have $\lvert \srptalg(J) \rvert = \Omega (\lvert \{\text{leaves of $\mathcal{S}'$}\}\rvert)$. Thus, it remains to show the following:
    
    \begin{lem}\label{lem_srpt_resp}
        For $\alpha = O(1)$ sufficiently large, $\lvert \srptalg(J) \rvert = \Omega(\lvert \mathcal{S} \rvert)$
    \end{lem}
    \begin{proof}
        We first show that for the majority of jobs $j$ in $\mathcal{S}$'s forest, we run $j$ itself on some machine for at least a constant fraction of the time interval $[r_j, r_j + \frac{\ell_j}{2}]$.
        
        \begin{prop}\label{prop_find_interval}
            For at least half of the nodes $j$ in $\mathcal{S}$'s forest, there exists a closed interval $I_j \subset [r_j, r_j + \frac{\ell_j}{2}]$ of length at least $\frac{\ell_j}{8}$ such that $\mathcal{S}$ runs $j$ on some machine during $I_j$.
        \end{prop}
        \begin{proof}
            We say a node $j$ is a \emph{non-progenitor} if $j$ has less than $2^z$ descendants at depth $z$ from $j$ for all $z \geq 1$. Because $\mathcal{S}$ satisfies (\ref{prop_resolute}), at least half of the nodes in $\mathcal{S}$'s forest are non-progenitors. This follows from Lemma 7 in \cite{KalyanasundaramP03}.
            
            Now consider any non-progenitor node $j$. Because $\mathcal{S}$ is a forest, $\mathcal{S}$ is only running $j$ or its descendants on some machine in times $[r_j, r_j + \frac{\ell_j}{2}]$. Further, because $j$ is a non-progenitor and $\mathcal{S}$ satisfies (\ref{prop_resolute}) and (\ref{prop_shrinking}), we can partition $[r_j, r_j + \frac{\ell_j}{2}] = [r_j, a] \cup (a,b) \cup [b, r_j + \frac{\ell_j}{2}]$ such that $[r_j, a]$ and $[b, r_j + \frac{\ell_j}{2}]$ are times where $\mathcal{S}$ is running $j$, and $(a,b)$ are times where $\mathcal{S}$ is running descendants of $j$. By taking $\alpha$ sufficiently large, we have $\lvert (a,b) \rvert \leq \frac{\ell_j}{4}$. This follows from Lemma 6 in \cite{KalyanasundaramP03}.
            It follows, at least one of $[r_j, a]$ or $[b, r_j + \frac{\ell_j}{2}]$ has length at least $\frac{\ell_j}{8}$. This gives the desired $I_j$. 
        \end{proof}
        
        Let $\mathcal{S}'' \subset \mathcal{S}$ be the collection of jobs guaranteed by the proposition, so $\lvert \mathcal{S}'' \rvert \geq \frac{1}{2} \lvert \mathcal{S} \rvert$. It suffices to show that $\lvert \srptalg(J) \rvert = \Omega(\lvert \mathcal{S}'' \rvert)$. Thus, we argue about $\mlaxalg(J)$ in the interval $I_j$ (guaranteed by \Cref{prop_find_interval}) for some $j \in \mathcal{S}''$.
        
        \begin{prop}\label{find_srpt_resp}
            Consider any job $j \in \mathcal{S}''$. For sufficiently large $\alpha = O(1)$, $\mlaxalg(J)$ starts running at least $\frac{m}{16}$ jobs during $I_j$ such that each such job $j'$ satisfies $[f_{j'}(\mlaxalg(J)), f_{j'}(\mlaxalg(J)) + x_{j'}] \subset I_j$.
        \end{prop}
        \begin{proof}
            We let $I'$ be the prefix of $I_j$ with length exactly $\frac{\lvert I_j \rvert}{2} \geq \frac{\ell_j}{16}$. Recall that $j \in \mathcal{S}''$ is non-viable. Thus, because $I' \subset I_j \subset [r_j, r_j + \frac{\ell_j}{2}]$, $\mlaxalg(J)$ is always running at least $\frac{1}{8}m$ jobs $j'$ satisfying $\alpha x_{j'} < \ell_j$ during $I'$.
            
            We define $J'$ to be the set of jobs that $\mlaxalg(J)$ runs during $I'$ satisfying $\alpha x_{j'} < \ell_j$. We further partition $J'$ into size classes, $J' = \bigcup_{z \in \mathbb{N}} J'_z$ such that $J'_z$ consists of the jobs in $J'$ with size in $(\frac{\ell_j}{\alpha^{z+1}}, \frac{\ell_j}{\alpha^z}]$.
            
            For each machine $i$, we let $T_z^i$ be the times in $I'$ that $\mlaxalg(J)$ is running a job from $J_z'$ on machine $i$. Note that each $T_z^i$ is the union of finitely many intervals. Then because $\mlaxalg(J)$ is always running at least $\frac{1}{8}m$ jobs $j'$ satisfying $\alpha x_{j'} \leq \ell_j$ during $I'$, we have:
            \[\sum_{z \in \mathbb{N}} \sum_{i \in [m]} \lvert T_z^i \rvert \geq \frac{m}{8} \lvert I' \rvert.\]
            By averaging, there exists some $z$ with $\sum_{i \in [m]} \lvert T_z^i \rvert \geq \frac{m}{8} \frac{\lvert I' \rvert}{2^{z+1}}$.
            
            Fix such a $z$. It suffices to show that there exist at least $\frac{m}{16}$ jobs in $J_z'$ that \laxalg~ starts in $I'$. This is because every job $j' \in J_z'$ has size at most $\frac{\ell_j}{\alpha^z}$ and $\lvert I_j \setminus I' \rvert \geq \frac{\ell_j}{16}$. Taking $\alpha \geq 16$ gives that $f_{j'}(\laxalg) + x_{j'} \in I_j$.
            
            Note that every job in $J'_z$ has size within a $\alpha$-factor of each other, so there can be at most one such job per stack at any time. This implies that there are at most $m$ jobs in $J'_z$ that \emph{don't} start in $I'$ (i.e. the start before $I'$.) These jobs contribute at most $m \frac{\ell_j}{\alpha^z}$ to $\sum_{i \in [m]} \lvert T_z^i \rvert$. Choosing $\alpha$ large enough, we can ensure that the jobs in $J'_z$ that start in $I'$ contribute at least $\frac{m}{16} \frac{\ell_j}{\alpha^z}$ to $\sum_{i \in [m]} \lvert T_z^i \rvert$. To conclude, we note that each job in $J_z'$ that starts in $I'$ contributes at most $\frac{\ell_j}{\alpha^z}$ to the same sum, so there must exist at least $\frac{m}{16}$ such jobs. 
        \end{proof}
        
        Using the above proposition, we define a charging scheme to show that $\lvert \srptalg(J) \rvert = \Omega(\lvert \mathcal{S}'' \rvert)$. Each job $j \in \mathcal{S}''$ begins with $1$ credit. By the proposition, we can find $\frac{m}{16}$ jobs $j'$ such that $[f_{j'}(\mlaxalg(J)), f_{j'}(\mlaxalg(J)) + x_{j'}]$ is contained in the time when $\mathcal{S}$ runs $j$. There are two cases to consider. If all $\frac{m}{16}$ jobs we find are contained in $\srptalg(J)$, then we transfer $\frac{16}{m}$ credits from $j$ to each of the $\frac{m}{16}$-many jobs. Note that here we are using $m \geq 16$. Otherwise, there exists some such $j'$ that is not in $\srptalg(J)$. Then $ \srptalg(J)$ will complete at least $m$ jobs in $[f_{j'}(\mlaxalg(J)), f_{j'}(\mlaxalg(J)) + x_{j'}]$. We transfer $\frac{1}{m}$ credits from $j$ to each of the $m$-many jobs. To conclude, we note that each $j' \in \srptalg(J)$ gets $O(\frac{1}{m})$ credits from at most $O(m)$ jobs in $\mathcal{S}''$.
    \end{proof}
    
\end{proof}

\section{\srptalg and \mlaxalg are Competitive with Viable Jobs}\label{sec_viable}

We have shown that \srptalg is competitive with the non-viable, low-laxity jobs. Thus, it remains to account for the viable, low-laxity jobs. We recall that a job $j$ is viable if there exists a time in $[r_j, r_j + \frac{\ell_j}{2}]$ such that there are at least $\frac{7}{8}m$ jobs $j'$ on the frontier satisfying $\alpha x_{j'} \geq \ell_j$. The first such time is the pseudo-release time, $\tilde{r}_j$ of job $j$. For these jobs, we show that \srptalg and \mlaxalg together are competitive with the viable, low-laxity jobs of the optimal schedule.

\begin{thm}\label{thm_viable}
    Let $J$ be a set of jobs satisfying $\ell_j \leq x_j$ for all $j \in J$. Then for $\alpha = O(1)$ sufficiently large and number of machines $m \geq 8$, we have $\lvert \srptalg(J) \rvert + \lvert \mlaxalg(J) \rvert = \Omega(\lvert \Opt(J_v) \rvert)$, where $J_v$ is the set of viable jobs with respect to $\mlaxalg(J)$.
\end{thm}
\begin{proof}[Proof of \Cref{thm_viable}]
    Let $\mathcal{S}, \mathcal{S}'$ be the schedules guaranteed by \Cref{lem_opt_struct} on the set of jobs $J_v$ with $\hat{r}_j = \tilde{r}_j$ for all $j \in J_v$. We re-state the properties of these schedules for convenience:
    \begin{enumerate}
        \item Both $\mathcal{S}$ and $\mathcal{S}'$ complete every job they run.
        \item Let $J_i$ be the set of jobs that $\mathcal{S}$ runs on machine $i$. For every machine $i$ and time, if there exists a feasible job in $J_i$, then $\mathcal{S}$ runs such a job.
        \item For all jobs $j \in \mathcal{S}$, we have $f_j(\mathcal{S}) = \tilde{r}_j$.
        \item If job $j'$ is a descendant of job $j$ in $\mathcal{S}$, then $\alpha x_{j'} \leq \ell_j$
        \item $\lvert \{\text{leaves of $\mathcal{S}'$}\}\rvert + \lvert \mathcal{S} \rvert = \Omega(\lvert \Opt(J_v) \rvert)$.
    \end{enumerate}
    By \Cref{lem_srpt_leaves}, we have $\lvert \srptalg(J) \rvert = \Omega (\lvert \{\text{leaves of $\mathcal{S}'$}\}\rvert)$. Thus, it suffices to show that $\lvert \srptalg(J) \rvert + \lvert \mlaxalg(J) \rvert = \Omega(\lvert \mathcal{S} \rvert)$. We do this with two lemmas, whose proofs we defer until later. First, we show that \mlaxalg pushes (not necessarily completes) many jobs. In particular, we show:
    
    \begin{lem}\label{lem_lax_push}
        $\lvert \srptalg(J) \rvert + \#(\text{pushes of $\mlaxalg(J)$}) = \Omega(\lvert \mathcal{S} \rvert)$
    \end{lem}
    
    The main idea to prove \Cref{lem_lax_push} is to consider sequences of preemptions in \Opt. In particular, suppose \Opt preempts job $a$ for $b$ and then $b$ for $c$. Roughly, we use viability to show that the only way \mlaxalg doesn't push any of these jobs is if in between their pseudo-release times, \mlaxalg pushes $\Omega(m)$ jobs.
    
    Second, we show that the pushes of \mlaxalg give a witness that \srptalg and \mlaxalg together actually complete many jobs.
    
    \begin{lem}\label{lem_lax_srpt_complete}
        $\lvert \srptalg(J) \rvert + \lvert \mlaxalg(J) \rvert = \Omega(\#(\text{pushes of $\mlaxalg(J)$}))$.
    \end{lem}
    
    The main idea to prove \Cref{lem_lax_srpt_complete} is to upper-bound the number of jobs that \mlaxalg pops because they are infeasible (all other pushes lead to completed jobs.) The reason \mlaxalg pops a job $j$ for being infeasible is because while $j$ was on a stack, \mlaxalg spent at least $\frac{\ell_j}{2}$ units of time running jobs higher than $j$ on $j$'s stack. Either those jobs are completed by \mlaxalg, or \mlaxalg must have have done many pushes or replacements instead. We show that the replacements give a witness that \srptalg must complete many jobs.
    
    Combining these two lemmas completes the proof of \Cref{thm_viable}.
\end{proof}

Now we go back and prove \Cref{lem_lax_push} and \Cref{lem_lax_srpt_complete}.

\subsection{Proof of \Cref{lem_lax_push}}

Recall that $\mathcal{S}$ is a forest schedule. We say the \emph{first child} of a job $j$ is the child $j'$ of $j$ with the earliest starting time $f_{j'}(\mathcal{S})$. In other words, if $j$ is not a leaf, then its first child is the first job that pre-empts $j$. We first focus on a sequence of first children in $\mathcal{S}$.

\begin{lem}\label{lem_lax_single_push}
Let $a,b,c \in \mathcal{S}$ be jobs such that $b$ is the first child of $a$ and $c$ is the first child of $b$. Then $\mlaxalg(J)$ does at least one of the following during the time interval $[\tilde{r}_a, \tilde{r}_c]$:
\begin{itemize}
    \item Push at least $\frac{m}{8}$ jobs
    \item Push job $b$
    \item Push a job on top of $b$ when $b$ is on the frontier
    \item Push $c$
\end{itemize}
\end{lem}
\begin{proof}
    By the properties of $\mathcal{S}$, we have $\tilde{r}_a < \tilde{r}_b < \tilde{r}_c$. It suffices to show that if during $[\tilde{r}_a, \tilde{r}_c]$, $\mlaxalg(J)$ pushes strictly fewer than $\frac{m}{8}$ jobs, $\mlaxalg(J)$ does not push $b$, and $\mlaxalg(J)$ does not push any job on top of $b$ if $b$ is on the frontier, then $\mlaxalg(J)$ pushes $c$.
    
    First, because $\mlaxalg(J)$ pushes strictly fewer than $\frac{m}{8}$ jobs during $[\tilde{r}_a, \tilde{r}_c]$, there exists at least $\frac{7}{8} m$ stacks that receive no push during this interval. We call such stacks \emph{stable}. The key property of stable stacks is that the laxities of their top- and second-top jobs never decrease during this interval, because these stacks are only changed by replacements and pops.
    
    Now consider time $\tilde{r}_a$. By definition of pseudo-release time, at this time, there exist at least $\frac{7}{8} m$ stacks whose top job $j'$ satisfies $\alpha x_{j'} \geq \ell_j$. Further, for any such stack, let $j''$ be its second-top job. Then because $b$ is a descendant of $a$ in $\mathcal{S}$, we have:
    \[\alpha x_b \leq \ell_a \leq \alpha x_{j'} \leq \ell_{j''}.\]
    It follows, there exist at least $\frac{3}{4}m$ stable stacks whose second-top job $j''$ satisfies $\alpha x_b \leq \ell_{j''}$ for the entirety of $[\tilde{r}_a, \tilde{r}_c]$. We say such stacks are $b$-stable.
    
    Now consider time $\tilde{r}_b$. We may assume $b$ is not pushed at this time. However, there exist at least $\frac{3}{4}m$ that are $b$-stable. Thus, if we do not replace the top of some stack with $b$, it must be the case that the top job $j'$ of every $b$-stable stack satisfies $\ell_j' \geq \ell_b$. Because these stacks are stable, their laxities only increase by time $\tilde{r}_c$, so $\mlaxalg(J)$ will push $c$ on some stack at that time.
    
    Otherwise, suppose we replace the top job of some stack with $b$. In particular, $b$ is on the frontier at $\tilde{r}_b$. We may assume that no job is pushed directly on top of $b$. If $b$ remains on the frontier by time $\tilde{r}_c$, then $\mlaxalg(J)$ will push $c$, because $\alpha x_c \leq \ell_b$. The remaining case is if $b$ leaves the frontier in some time in $[\tilde{r}_b, \tilde{r}_c]$. We claim that it cannot be the case that $b$ is popped, because by (\ref{prop_efficient}), $\mathcal{S}$ could not complete $b$ by time $\tilde{r}_c$, so $\mlaxalg(J)$ cannot as well. Thus, it must be the case that $b$ is replaced by some job, say $d$ at time $\tilde{r}_d$. At this time, there exist at least $\frac{3}{4}m$ stacks whose second-top job $j''$ satisfies $\alpha x_d \leq \ell_{j''}$. It follows, there exist at least $\frac{m}{2}$ $b$-stable stacks whose second-top job $j''$ satisfies $\alpha x_d \leq \ell_{j''}$ at time $\tilde{r}_d$. Note that because $m \geq 8$, there exists at least one such stack, say $i$, that is not $b$'s stack. In particular, because $b$'s stack has minimum laxity, it must be the case that the top job $j'$ of stack $i$ satisfies $\ell_{j'} \geq \ell_b$. Finally, because stack $i$ is stable, at time $\tilde{r}_c$ we will push $c$.  
\end{proof}

Now using the above lemma, we give a charging scheme to prove \Cref{lem_lax_push}. First note that by \Cref{lem_srpt_leaves}, we have $\lvert \srptalg(J) \rvert = \Omega(\#(\text{leaves of $\mathcal{S}$}))$. Thus, it suffices to give a charging scheme such that each job $a \in \mathcal{S}$ begins with $1$ credit, and charges it to leaves of $\mathcal{S}$ and completions of $\mlaxalg(J)$ so that each job is charged at most $O(1)$ credits. Each job $a \in \mathcal{S}$ distributes its $1$ credit as follows:

    \begin{itemize}
        \item (Leaf Transfer) If $a$ is a leaf or parent of a leaf of $\mathcal{S}$, say $\ell$, then $a$ charges $\ell$ for $1$ credit.
    \end{itemize}
        Else let $b$ be the first child of $a$ and $c$ the first child of $b$ in $\mathcal{S}$
        \begin{itemize}
            \item (Push Transfer)  If $\mlaxalg(J)$ pushes $b$ or $c$, then $a$ charges 1 unit to $b$ or $c$, respectively.
            \item (Interior Transfer) Else if job $b$ is on the frontier, but another job, say $d$, is pushed on top of $b$, then $a$ charges $1$ unit to $d$.
            \item ($m$-Push Transfer) Otherwise, by \Cref{lem_lax_single_push}, $\mlaxalg(J)$ must push at least $\frac{m}{8}$ jobs during $[\tilde{r}_a, \tilde{r}_c]$. In this case, $a$ charges $\frac{8}{m}$ units to each of these $\frac{m}{8}$ such jobs.
        \end{itemize}
    This completes the description of the charging scheme. It remains to show that each job is charged at most $O(1)$ credits. Each job receives at most $2$ credits due to Leaf Transfers and at most $2$ credits due to Push Transfers and Interior Transfers. As each job is in at most $3m$ intervals of the form $[\tilde{r}_a, \tilde{r}_c]$,  each job is charged $O(1)$ from $m$-Push Transfers.

\subsection{Proof of \Cref{lem_lax_srpt_complete}}

    Recall in \mlaxalg, there are two types of pops: a job is popped if it is completed, and then we continue popping until the top job of that stack is feasible. We call the former \emph{completion pops} and the later \emph{infeasible pops}. Note that it suffices to prove the next lemma, which bounds the infeasible pops. This is because $\#(\text{pushes of $\mlaxalg(J)$}) = \#(\text{completions pops of $\mlaxalg(J)$}) + \#(\text{infeasible pops of $\mlaxalg(J)$})$. To see this, note that every stack is empty at the beginning and end of the algorithm, and the stack size only changes due to pushes and pops. 
    
    \begin{lem}\label{lem_lax_srpt_pop}
        For $\alpha = O(1)$ sufficiently large, we have: \[\lvert \srptalg(J) \rvert + \lvert \mlaxalg(J) \rvert + \#(\text{pushes of $\mlaxalg(J)$}) \geq 2 \cdot \#(\text{infeasible pops of $\mlaxalg(J)$}).\]
    \end{lem}
        \begin{proof}
        We define a charging scheme such that the completions of $\srptalg(J)$ and $\mlaxalg(J)$ and the pushes executed by $\mlaxalg(J)$  pay for the infeasible pops. Each completion of $\srptalg(J)$ is given $2$ credits, each completion of $\mlaxalg(J)$ is given $1$ credit, and each job that $\mlaxalg(J)$ pushes is given $1$ credit. Thus each job begins with at most $4$ credits. For any $z \geq 0$, we say job $j'$ is $z$-below $j$ (at time $t$) if $j'$ and $j$ are on the same stack in $\mlaxalg(J)$ and $j'$ is $z$ positions below $j$ on that stack at time $t$. We define $z$-above analogously. A job $j$ distributes these initial credits as follows:
        \begin{itemize}
            \item(\srptalg-transfer) If $\srptalg(J)$ completes job $j$ and \mlaxalg~ also ran $j$ at some point, then $j$ gives $\frac{1}{2^{z+1}}$ credits to the job that is $z$-below $j$ at time $f_j(\mlaxalg(J))$ for all $z \geq 0$.
            \item($m$-\srptalg-transfer) If $\srptalg(J)$ completes job $j$ at time $t$, then $j$ gives $\frac{1}{2^{z+1}} \frac{1}{m}$ credits to the job that is $z$-below the top of each stack in $\mlaxalg(J)$ at time $t$ for all $z \geq 0$.
            \item(\mlaxalg-transfer) If $\mlaxalg(J)$ completes a job $j$, then $j$ gives $\frac{1}{2^{z+1}}$ credits to the job that is $z$-below $j$ at the time $j$ is completed for all $z \geq 0$.
            \item(Push-transfer) If $\mlaxalg(J)$ pushes a job $j$, then $j$ gives $\frac{1}{2^{z+1}}$ credits to the job that is $z$-below $j$ at the time $j$ is pushed for all $z \geq 0$.
        \end{itemize}
        
        It remains to show that for $\alpha = O(1)$ sufficiently large, every infeasible pop gets at least $4$ credits. We consider any job $j$ that is an infeasible pop of $\mlaxalg(J)$. At time $\tilde{r}_j$ when $j$ joins some stack in $\mlaxalg(J)$, say $H$, $j$'s remaining laxity was at least $\frac{\ell_j}{2}$. However, as $j$ later became an infeasible pop, it must be the case that while $j$ was on stack $H$, $\mlaxalg(J)$ was running jobs that are higher than $j$ on stack $H$ for at least $\frac{\ell_j}{2}$ units of time.
        
        Let $I$ be the union of intervals of times that $\mlaxalg(J)$ runs a job higher than $j$ on stack $H$ (so $j$ is on the stack for the entirety of $I$.) Then we have $\lvert I \rvert \geq \frac{\ell_j}{2}$. Further, we partition $I$ based on the height of the job 
        on $H$ that $\mlaxalg(J)$ is currently running. In particular,
        we partition $I = \bigcup_{z \geq 1} I_z$, where $I_z$ is the union of intervals of times that $\mlaxalg(J)$ runs a job on $H$ that is exactly $z$-above $j$.
        
        By averaging, there exists a $z \geq 1$ such that $\lvert I_z \rvert \geq \frac{\ell_j}{2^{z+1}}$. Fix such a $z$. We can write $I_z$ as the union of disjoint intervals, say $I_z = \bigcup_{u = 1}^s [a_u, b_u]$.
        Because during each sub-interval, $\mlaxalg(J)$ is running jobs on $H$
        that are much smaller than $j$ itself, these jobs give a witness that $\srptalg(J)$ completes many jobs as long as these sub-intervals are long enough.
        We formalize this in the following proposition.
        
        \begin{prop}\label{prop_subinterval_srpt}
            In each sub-interval $[a_u,b_u]$ of length at least $4\frac{\ell_j}{\alpha^z}$, job $j$ earns at least $\frac{1}{2^{z+3}} \frac{b_u - a_u}{\ell_j/\alpha^z}$ credits from \srptalg-transfers and $m$-\srptalg-transfers.
        \end{prop}
        \begin{proof}
            Because $[a_u,b_u]$ has length at least $4\frac{\ell_j}{\alpha^z}$, we can partition $[a_u,b_u]$ into sub-sub-intervals such that all but at most one sub-sub-interval has length exactly $2 \frac{\ell_j}{\alpha^z}$. In particular, we have at least $\frac{1}{2} \frac{b_u - a_u}{2 \ell_j / \alpha^z}$ sub-sub intervals of length exactly $2 \frac{\ell_j}{\alpha^z}$.
            
            Now consider any such sub-sub-interval. During this time, $\mlaxalg(J)$ only runs jobs on $H$ that are $z$-above  $j$. Let $J_z$ be the set of $z$-above jobs that $\mlaxalg(J)$ runs during $I_z$. For every job $j' \in J_z$, we have $x_{j'} \leq \frac{\ell_j}{\alpha^z}$. It follows that $j'$ is on  stack $H$ for at most $x_{j'} \leq \frac{\ell_j}{\alpha^z}$ units of time. In particular, $\mlaxalg(J)$ must \emph{start} a new $z$-above job, say $j'$, in the first half of the sub-sub-interval at some time, say $t$.
            
            At time $t$, $j'$ is feasible. There are two cases to consider. If $\srptalg(J)$ also completes $j'$ at some time, then $j$ get $\frac{1}{2^{z+1}}$ credits from $j'$ in a \srptalg-transfer. Otherwise if $\srptalg(J)$ never completes $j'$, then  because $j'$ is feasible at $t$, it must be the case that $\srptalg(J)$ completes $m$ jobs during the sub-sub-interval. Thus, $j$ gets $\frac{1}{m} \frac{1}{2^{z+1}}$ credits from $m$ separate $m$-\srptalg-transfers during this sub-sub-interval. We conclude, job $j$ gets at least $\frac{1}{2^{z+1}}$ credits from at least $\frac{1}{2} \frac{b_u - a_u}{2 \ell_j / \alpha^z}$ sub-sub-intervals.          
        \end{proof}
        
        On the other hand, even if the sub-intervals are too short, the job $j$ still gets credits from \mlaxalg-transfers and Push-transfers when the height of the stack changes.
        We formalize this in the following proposition.
        
        \begin{prop}\label{prop_subinterval_lax_push}
            For every sub-interval $[a_u,b_u]$, job $j$ earns at least $\frac{1}{2^{z+2}}$ credits from \mlaxalg-transfers and Push-transfers at time $b_u$.
        \end{prop}
        \begin{proof}
            Up until time $b_u$, $\mlaxalg(J)$ was running a $z$-above job on  stack $H$. At time $b_u$, the height of the stack $H$ must change. If the height decreases, then it must be the case that $\mlaxalg(J)$ completes the $z$-above job, so $j$ will get $\frac{1}{2^{z+1}}$ credits from a $\mlaxalg(J)$-transfer. Otherwise, the height increases, so $\mlaxalg(J)$~ must push a job that is $z+1$-above $j$, which gives $j$ $\frac{1}{2^{z+2}}$ credits from a Push-transfer. 
        \end{proof}
        
        Now we combine the above two propositions to complete the proof
        of Lemma \ref{lem_lax_srpt_pop}. We say a sub-interval $[a_u,b_u]$ is \emph{long} if it has length at least $4\frac{\ell_j}{\alpha^z}$ (i.e. we can apply \Cref{prop_subinterval_srpt} to it) and \emph{short} otherwise. First, suppose the aggregate length of all long intervals it at least $4 \cdot 2^{z+3} \frac{\ell_j}{\alpha^z}$. Then by \Cref{prop_subinterval_srpt}, job $j$ gets at least $4$ credits from the long intervals. Otherwise, the aggregate length of all long intervals is less than $4 \cdot 2^{z+3} \frac{\ell_j}{\alpha^z}$. In this case, recall that the long and short intervals partition $I_z$, which has length at least $\frac{\ell_j}{2^{z+1}}$. It follows, the aggregate length of the short intervals is at least $\frac{\ell_j}{2^{z+1}} - 4 \cdot 2^{z+3} \frac{\ell_j}{\alpha^z}$. For $\alpha = O(1)$ large enough, we may assume the aggregate length of the short intervals is at least $4 \cdot 2^{z+2} \frac{4\ell_j}{\alpha^z}$. Because each short interval has length at most $4\frac{\ell_j}{\alpha^z}$, there are at least $4 \cdot 2^{z+2}$ short intervals. We conclude, by \Cref{prop_subinterval_lax_push}, job $j$ gets at least $4$ credits from the short intervals. We conclude, in either case job $j$ gets at least $4$ credits.
    \end{proof}

\section{Putting it all together}\label{sec_final}

In this section, we prove our main result, \Cref{thm_main_deterministic}, which follows from the next meta-theorem:

\begin{thm}\label{thm_main_meta}
    Let $J$ be any set of jobs. Then for number of machines $m \geq 16$, we have $\lvert \LMNYalg(J_{hi}) \rvert + \lvert \srptalg(J_{lo}) \rvert + \lvert \mlaxalg(J_{lo}) \rvert = \Omega(\lvert \Opt(J) \rvert)$, where $J_{hi} = \{j \in J \mid \ell_j > x_j \}$ and $J_{lo} = \{j \in J \mid \ell_j \leq x_j\}$ partition $J$ into high- and low-laxity jobs.
\end{thm}
\begin{proof}
    We have $\lvert \LMNYalg(J_{hi}) \rvert = \Omega(\lvert \Opt(J_{hi}) \rvert$ by \Cref{lem_LMNY}. Also, we further partition $J_{lo} = J_v \cup J_{nv}$ into the viable and non-viable jobs with respect to $\mlaxalg(J_{lo})$. Then \Cref{thm_srpt_nv_low} and \Cref{thm_viable} together give $\lvert \srptalg(J_{lo}) \rvert + \lvert \mlaxalg(J_{lo}) \rvert = \Omega (\lvert \Opt(J_v) \rvert + \lvert \Opt(J_{nv}) \rvert)$. To complete the proof, we observe that $J = J_{hi} \cup J_v \cup J_{nv}$ partitions $J$, so $\lvert \Opt(J_{hi}) \rvert + \lvert \Opt(J_v) \rvert + \lvert \Opt(J_{nv}) \rvert = \Omega(\lvert \Opt(J) \rvert)$.
\end{proof}

The proof of \Cref{thm_main_final}, which gives our performance guarantee for \finalalg is immediate:

\begin{proof}[Proof of \Cref{thm_main_final}]
    By combining \Cref{thm_main_meta} and \Cref{lem_reduce_machine}, the objective value achieved by \finalalg is:
    \begin{align*}
        \Omega(\lvert \LMNYalg(J_{hi}, \frac{m}{3})\rvert + \lvert \srptalg(J_{lo}, \frac{m}{3})\rvert + \lvert \mlaxalg(J_{lo}, \frac{m}{3})\rvert) &= \Omega( \lvert \Opt(J, \frac{m}{3}) \rvert)\\
        &= \Omega(\lvert \Opt(J,m) \rvert). 
    \end{align*}
\end{proof}

Finally, we obtain our $O(1)$-competitive deterministic algorithm for all $m > 1$ (recall \finalalg is $O(1)$-competitive only when $m \geq 48$) by using a two-machine algorithm when $m$ is too small:

\begin{proof}[Proof of \Cref{thm_main_deterministic}]
    Our algorithm is the following: If $1 < m < 48$, then we run the deterministic two-machine algorithm from \cite{KalyanasundaramP03} which is $O(1)$-competitive with the optimal single-machine schedule. Thus by \Cref{lem_reduce_machine}, this algorithm is also $O(m) = O(1)$-competitive for all $m < 48$. Otherwise, $m \geq 48$, so we run \finalalg.
\end{proof}

\bibliographystyle{alpha}

\bibliography{references}

\newcommand{\etalchar}[1]{$^{#1}$}
\begin{thebibliography}{BKM{\etalchar{+}}92}

\bibitem[AC18]{AzarC18}
Yossi Azar and Sarel Cohen.
\newblock An improved algorithm for online machine minimization.
\newblock {\em Operations Research Letters}, 46(1):128--133, 2018.

\bibitem[BKM{\etalchar{+}}92]{BaruahKMMRRSW92}
Sanjoy~K. Baruah, Gilad Koren, Decao Mao, Bhubaneswar Mishra, Arvind
  Raghunathan, Louis~E. Rosier, Dennis~E. Shasha, and Fuxing Wang.
\newblock On the competitiveness of on-line real-time task scheduling.
\newblock {\em Real Time Systems}, 4(2):125--144, 1992.

\bibitem[CEM{\etalchar{+}}20]{ChenEMSS20}
Lin Chen, Franziska Eberle, Nicole Megow, Kevin Schewior, and Clifford Stein.
\newblock A general framework for handling commitment in online throughput
  maximization.
\newblock {\em Mathematical Programming}, 183(1):215--247, 2020.

\bibitem[CMS18]{CMS18}
Lin Chen, Nicole Megow, and Kevin Schewior.
\newblock An o(log m)-competitive algorithm for online machine minimization.
\newblock {\em {SIAM} Journal of Computing}, 47(6):2057--2077, 2018.

\bibitem[EMS20]{EberleMS20}
Franziska Eberle, Nicole Megow, and Kevin Schewior.
\newblock Optimally handling commitment issues in online throughput
  maximization.
\newblock In Fabrizio Grandoni, Grzegorz Herman, and Peter Sanders, editors,
  {\em European Symposium on Algorithms)}, volume 173 of {\em LIPIcs}, pages
  41:1--41:15. Schloss Dagstuhl - Leibniz-Zentrum f{\"{u}}r Informatik, 2020.

\bibitem[IMPS17]{ImMPS17}
Sungjin Im, Benjamin Moseley, Kirk Pruhs, and Clifford Stein.
\newblock An o(log log m)-competitive algorithm for online machine
  minimization.
\newblock In {\em 2017 {IEEE} Real-Time Systems Symposium}, pages 343--350.
  {IEEE} Computer Society, 2017.

\bibitem[KP00]{KalyanasundaramP00}
Bala Kalyanasundaram and Kirk Pruhs.
\newblock Speed is as powerful as clairvoyance.
\newblock {\em Journal of the {ACM}}, 47(4):617--643, 2000.
\newblock Also 1995 Symposium on Foundations of Computer Science.

\bibitem[KP01]{KalyanasundaramP01}
Bala Kalyanasundaram and Kirk Pruhs.
\newblock Eliminating migration in multi-processor scheduling.
\newblock {\em Journal of Algorithms}, 38(1):2--24, 2001.

\bibitem[KP03]{KalyanasundaramP03}
Bala Kalyanasundaram and Kirk Pruhs.
\newblock Maximizing job completions online.
\newblock {\em Journal of Algorithms}, 49(1):63--85, 2003.
\newblock Also 1998 European Symposium on Algorithms.

\bibitem[KS94]{KorenS94}
Gilad Koren and Dennis~E. Shasha.
\newblock {MOCA:} {A} multiprocessor on-line competitive algorithm for
  real-time system scheduling.
\newblock {\em Theoretical Computer Science}, 128(1{\&}2):75--97, 1994.

\bibitem[LMNY13]{LucierMNY13}
Brendan Lucier, Ishai Menache, Joseph Naor, and Jonathan Yaniv.
\newblock Efficient online scheduling for deadline-sensitive jobs.
\newblock In Guy~E. Blelloch and Berthold V{\"{o}}cking, editors, {\em {ACM}
  Symposium on Parallelism in Algorithms and Architectures}, pages 305--314.
  {ACM}, 2013.

\bibitem[PSTW02]{PhillipsSTW02}
Cynthia~A. Phillips, Clifford Stein, Eric Torng, and Joel Wein.
\newblock Optimal time-critical scheduling via resource augmentation.
\newblock {\em Algorithmica}, 32(2):163--200, 2002.

\end{thebibliography}

\end{document}